\documentclass[twocolumn,showpacs,pre,preprintnumbers,amsmath,amssymb]{revtex4}
\usepackage{epsfig}
\usepackage{graphicx}
\usepackage{dcolumn}
\usepackage{bm}
\begin{document}
\title{Networking Effects on Cooperation in Evolutionary Snowdrift Game}
\author{Li-Xin Zhong}
\author{Da-Fang Zheng}\email{dfzheng@zjuem.zju.edu.cn}
\author{B. Zheng}
\affiliation{ Zhejiang Institute of Modern Physics and Department
of Physics, Zhejiang University, Hangzhou 310027, People's
Republic of China}
\author{Chen Xu}
\author{P. M. Hui}
\affiliation{Department of Physics and Institute of
Theoretical Physics, The Chinese University of Hong Kong, Shatin,
Hong Kong, China}

\date{\today}

\begin{abstract}
The effects of networking on the extent of cooperation emerging in
a competitive setting are studied.  The evolutionary snowdrift
game, which represents a realistic alternative to the well-known
Prisoner's Dilemma, is studied in the Watts-Strogatz network that
spans the regular, small-world, and random networks through random
re-wiring.  Over a wide range of payoffs, a re-wired network is
found to suppress cooperation when compared with a well-mixed or
fully connected system. Two extinction payoffs, that characterize
the emergence of a homogeneous steady state, are identified.  It
is found that, unlike in the Prisoner's Dilemma, the standard
deviation of the degree distribution is the dominant property in a
re-wired network that governs the extinction payoffs.

\end{abstract}


\maketitle

\section{Introduction}

Evolutionary game theory has become an important tool for
investigating and understanding cooperative or altruistic behavior
in systems consisting of competitive entities.  These systems may
occur in a wide range of problems in different disciplines and
include physical, biological, ecological, social, and political
systems.  Mathematicians, biologists, and physicists alike have
found the phenomena of emergence of cooperative behavior
fascinating.  Since the ground-breaking work on repeated or
iterated games based on the Prisoner's Dilemma (PD) game by
Axelrod \cite{axelrod,axelrod1}, there has been a continuous
effort on exploring the determining factors on possible
cooperative behavior in evolutionary games based on the PD game
and its variations
\cite{trivers,nowak,nowak1,hauert,nowak2,lieberman}, with a recent
emphasis on the effects of spatial structures such as regular
lattices \cite{nowak1,hauert,nowak3,doebeli,killingback} and
networks \cite{lieberman,abramson,kim,ebel,masuda,wu,santos}.
Remarkably, it was found that cooperation can be induced in a
repeated PD game by cleverly designed strategies.  Spatial
structures, e.g., lattices or networks, are found to favor
cooperative behavior in the evolutionary PD game.  The two
subjects involved in networked games, i.e., emergence phenomenon
and the physics of networks \cite{albert}, are among the most
rapidly growing branches in physics.

The present work was motivated by the recent concerns on whether
the PD game should be the sole model for studying emerging
cooperative phenomena \cite{hauert}.  Due to practical
difficulties in accurately quantifying the payoffs in game theory,
the snowdrift game (SG) has been proposed as possible alternative
to the PD game.  Previous work on the SG have focused on the
effects of connectivity in structures such as lattices
\cite{hauert} and fully connected networks \cite{hauert,hofbauer}.
The latter is also referred to as the well-mixed case in the
literature. Here, we investigate the networking effects on an
evolutionary snowdrift game \cite{hauert,sugden,smith} within the
Watts-Strogatz (WS) \cite{watts,albert} model of small world
constructed by randomly re-wiring a regular network. Starting with
a random mixture of competing nodes of opposite characters, we
found that (i) the steady-state population may consist of only one
kind of nodes or a mixture of nodes of different characters,
depending on degree $K$ in a regular lattice before re-wiring and
the extent of re-wiring $q$; (ii) for a wide range of payoffs, a
re-wired network {\em suppresses} the fraction of cooperative
nodes in the steady state and hence the overall level of
cooperation, when compared with a fully connected network or
well-mixed case; (iii) networking effect on the critical payoffs
for the extinction of one kind of nodes in the steady state
depends sensitively on the width or standard deviation of the
degree distribution induced by re-wiring.

The plan of the paper is as follows. The evolutionary snowdrift
game and the Watts-Strogatz network structure are introduced in
Sec. II.  Section III gives the numerical results in regular
lattices.  Two extinction payoffs are identified.  An analytic
expression for the extinction payoff of defective character is
given, together with a discussion on the discrepancy between
observed value and theoretical value of the extinction payoff for
runs with finite time steps.  Section IV gives the numerical
results in re-wired networks.  The dependence on the re-wiring
probability is found to be governed by the standard deviation of
the degree distribution of the re-wired network.  In Sec. V, we
give an approximate expression for the extinction payoff and
summarize our results.

\section{Model}

The basic snowdrift game, which is equivalent to the hawk-dove or
chicken game \cite{sugden,smith}, is most conveniently described
using the following scenario.  Consider two drivers hurrying home
in opposite directions on a road blocked by a snowdrift. Each
driver has two possible actions -- to shovel the snowdrift
(cooperate (C)) or not to do anything (not-to-cooperate or
``defect" (D)). This is similar to the PD game in which each
player has two options: to cooperate or to defect. If the two
drivers cooperate, they could be back home on time and each will
get a reward of $b$. Shovelling is a laborious job with a total
cost of $c$.  Thus, each driver gets a net reward of $R=b - c/2$.
If both drivers take action D, they both get stuck, and each gets
a reward of $P=0$.  If only one driver takes action C and shovels
the snowdrift, then both drivers can get through.  The driver
taking action D (not to shovel) gets home without doing anything
and hence gets a payoff $T=b$, while the driver taking action C
gets a ``sucker" payoff of $S=b-c$. The SG is, therefore, given by
the following payoff matrix:
\begin{equation}
\begin{array}{ccc}
\ & \begin{array}{cc} C & D \end{array} \\
\begin{array}{c} C \\ D \end{array}&\left(\begin{array}{cc}R&S \\
T&P\end{array}\right).
\end{array}
\end{equation}
The matrix element gives the payoff to a player using a strategy
listed in the left hand column when the opponent uses a strategy
in the top row.  The SG refers to the case $b>c>0$, leading to
$T>R>S>P$. This ordering of the payoffs {\em defines} the SG.
Without loss of generality, it is useful to assign $R=1$ so that
the payoffs can be characterized by a single parameter $r = c/2 =
c/(2b-c)$ for the cost-to-reward ratio. In terms of $0< r<1$, we
have $T=1+r$, $R=1$, $S=1-r$, and $P=0$.  For a player, the best
action is: to take D if the opponent takes C, otherwise take C.  A
larger value of $r$ tends to encourage the action D. The SG
becomes the PD when the cost $c$ is high such that $2b>c>b>0$,
which amounts to the ranking $T>R>P>S$ \cite{neumann,rapoport}.
Therefore, the SG and PD game differ only by the ordering of $P$
and $S$. Due to the difficulty in measuring payoffs in game
theory, the SD has been proposed to be a possible alternative to
the PD game in studying emerging cooperative phenomena
\cite{hauert,hofbauer}.

Evolutionary snowdrift game amounts to letting the character of a
connected population of inhomogeneous players evolve, according to
their performance \cite{hauert}. Consider $N$ players represented
by the nodes of a network.  Initially, the nodes are randomly
assigned to be either of C or D character.  The character of each
node is updated every time step simultaneously, according to the
following procedure.  At each time step, every node $i$ interacts
with all its connected $k_{i}$ neighbors and gets a payoff per
neighbor $\overline{V}_{i} = V_{i}/k_{i}$, where $V_{i}$ is
obtained by summing up the $k_{i}$ payoffs after comparing
characters with its neighbors.  Here, $k_{i}$ is the degree of
node $i$.  Every node $i$ then randomly selects a neighbor $j$ for
possible updating or evolution.  To compare the performance of the
two nodes $i$ and $j$, we construct
\begin{equation} w_{ij}=(\overline{V_{j}}-\overline{V_{i}})/(T-P) =
(\overline{V_{j}} - \overline{V_{i}})/(1+r).
\end{equation}
If $w_{ij} > 0$, then the character of node $i$ is {\em replaced}
by the character of node $j$ with probability $w_{ij}$, and thus
node $i$ becomes the offspring of node $j$.  The denominator
$(1+r)$ in $w_{ij}$ is, therefore, included as a normalization
factor. If $w_{ij} < 0$, then the character of node $i$ remains
unchanged.  In the long time limit, a state will be attained with
possible coexistence of nodes of both characters. The fraction of
$C$-nodes $f_C$, which is often called the frequency of
cooperators \cite{hauert}, measures the extent of cooperation in
the whole system and is determined by the structure of the
underlying network {\em and} the payoff parameter $r$.  In a fully
connected network, $f_C = 1-r$ \cite{hauert,hofbauer}. In
two-dimensional lattices with nearest-neighboring and next-nearest
neighboring connections, it has recently been observed that the
spatial structure tends to lower $f_C$ \cite{hauert}, when
compared with a fully connected network.  In contrast, spatial
connections are found to enhance $f_{C}$ in evolutionary PD games.
We note that one may define $w_{ij}$ in different ways. Popular
alternatives of $w_{ij}$ in evolutionary games include the use of
the total payoff instead of an average payoff per neighbor for
which the effects of a spread in the degrees of different nodes
will be much more prominent and allowing for the possibility of
replacing the character of a node by that of a neighboring node
with lower payoff.  In our choice of $w_{ij}$, a node will only
have a chance to be replaced if it encounters a better performing
node.

The physics of complex networks \cite{albert} is a fascinating
research area of current interest.  Here, the nodes in an
evolutionary SG are connected in the form of the Watts-Strogatz
\cite{watts} small-world model.  Starting with a one-dimensional
regular world consisting of a circular chain, i.e., periodic
boundary condition, of $N$ nodes with each node having a degree
$2K$ connecting to its $2K$ nearest neighbors, each of the $K$
links to the right hand side of a node is cut and re-wired to a
randomly chosen node with a probability $q$.  This simple model
gives the small world effect \cite{watts}, which refers to the
commonly observed phenomena of small separations between two
randomly chosen nodes in many real-life networks \cite{albert},
for $q \sim 0.1$. The parameter $q$ ($0 \leq q \ \leq 1$) thus
takes the network from a regular world through the small world to
a random world, with a fixed mean degree $\langle k \rangle = 2K$.
Here, we aim at understanding how the frequency of cooperators
$f_{C}$ behaves as the parameters characterizing the the spatial
structure $K$ and $q$ change, and as the payoff parameter $r$
characterizing the evolutionary dynamics changes.

\section{Results: Regular world}

\begin{figure}
\begin{center}
\epsfig{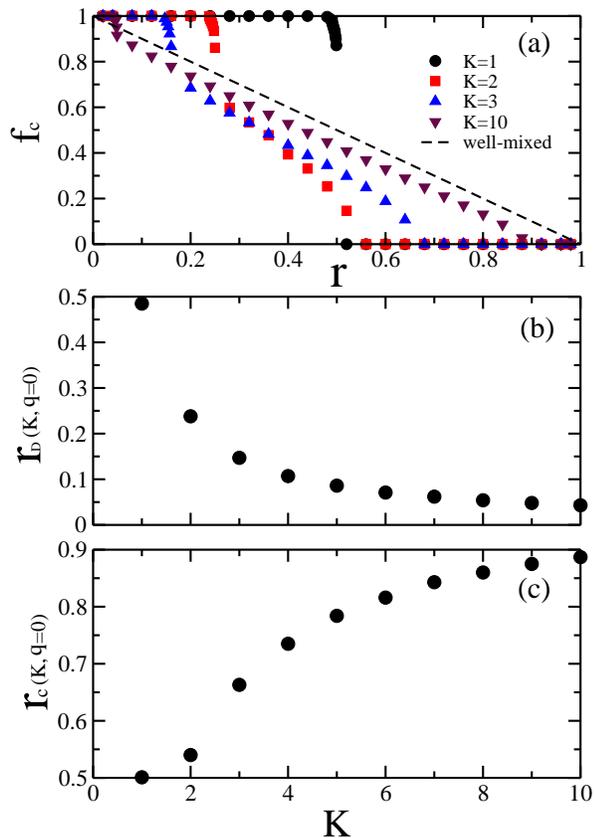} \caption{(Color
online) (a) The frequency of cooperators $f_{C}$ as a function of
the payoff parameter $r$ in a one-dimensional Watts-Strogatz
network of size $N=10^{3}$ nodes with $K=1$, $2$, $3$, $10$ before
re-wiring ($q=0$). The run time is $10^{4}$ time steps.  The
dashed line indicates the result $f_{C}=1-r$ for a fully connected
network. (b) The extinction payoff $r_{D}(K,q=0)$ as a function of
$K$. (c) The extinction payoff $r_{C}(K,q=0)$ as a function of
$K$.}
\end{center}
\end{figure}

Figure 1(a) shows $f_{C}$ as a function of $r$ in {\em regular
lattices} ($q=0$) of $N=10^{3}$ nodes with different values of
$K$.  The dashed line gives the result $f_{C}=1-r$ for the
well-mixed case.  The key features are: (i) there exists a value
$r_{D}(K,q=0)$ so that $f_{C}=1$ for $r < r_{D}$, i.e., the
extinction of D-nodes and all nodes become cooperative in nature,
(ii) for $r < r_{D}$, $f_{C}$ is enhanced when compared with the
well-mixed case, (iii) for a wide range of $r$, the frequency of
cooperators drops below that in the well-mixed case, (iv) there
exists a value $r_{C}(K,q=0)$ so that $f_{C} =0$ for $r
> r_{C}$, i.e., the extinction of cooperative nodes \cite{remark}.
These extinction payoffs thus characterize a transition between a
homogeneous and an inhomogeneous population as the payoff $r$ is
varied.  Figures 1(b) and 1(c) show that the extinction payoff for
defectors (cooperators) $r_{D}$ ($r_{C}$) decreases (increases)
monotonically with $K$ in regular lattices.  Numerically, it is
found that for $K=1,2$, $r_{D}(K,q=0)$ is close to $1/2K$; while
for $K \geq 3$, $r_{D}(K,q=0)$ is closer to $1/(2K+1)$.  In
addition, we observe that for $K \geq 3$, the relation $2r_{D} +
r_{C} \approx 1$ is satisfied.  The feature (iii) is analogous to
that observed in two-dimensional lattices \cite{hauert}.  We have
also found that $r_{D}$ and $r_{C}$ are independent of the number
of nodes $N$, as long as $N \gg 1$.  As the shortest path $L =
N/4K$ scales with $N$ in regular lattices, the results imply that
the $K$-dependence of the extinction payoffs does not come from
$L$, although $L$ is an important quantity in the description of
networks.  Clearly, the time it takes the system from the initial
configuration to that in the long time limit would depend on $N$
and hence $L$.

Analytically, the value of $r_{D}(K,q=0)$ can be estimated as
follows.  Recall that a higher value of $r$ promotes D-character
and $r_{D}$ refers to the highest value of $r$ below which no
D-node exists.  Since replacements are carried out
probabilistically, it is, therefore, important to consider cases
where there are only a few D-nodes and how these patterns will be
replaced. Note that for any value of $K$, a {\em single} D-node in
an otherwise C-node environment will {\em not} be replaced. The
reason is the following.  The average payoff for the single
D-node is $\overline{V_{D}} = 1+r$. Note that only the C-nodes
that are connected to the D-node may be involved in a possible
evolution. For these C-nodes, their average payoff
$\overline{V_{C(1D)}} = 1 - r/2K$, where the subscript indicate
that it is the average payoff for a C-node that is linked to one
D-node.  Thus, for the whole range of $r$, $\overline{V_{D}} >
V_{C(1D)}$ and thus a single D-node will not be replaced by its
connected C-neighbors.

This leads us to consider the patterns involving D-nodes, called
the last surviving patterns, that can be replaced by C-nodes in
one time step.  The case of $K=1$ serves to illustrate the key
ideas in estimating $r_{D}$.  For $K=1$, the single D-node
pattern will be evolved into one with a pair of neighboring
D-nodes.  The chain becomes that of ...CCCCDDCCCC....  Because
each D-node has a D-neighbor and the D-D payoff is $P=0$, the
average payoff of the D-nodes becomes $\overline{V_{D}} =
(1+r)/2$.  The payoff of the C-nodes that are linked to a D-node
is $\overline{V_{C(1D)}} = 1 - r/2$.  By equating
$\overline{V_{D}}$ and $\overline{V_{C(1D)}}$, we obtain an
estimate of $r_{D} = 1/2$ for $K=1$, which agrees with numerical
results.  For $r < 1/2$, $\overline{V_{C(1D)}} >
\overline{V_{D}}$ and thus the D-D dimer pattern may evolve into
an all-C chain, provided that both D-nodes randomly pick the
neighboring C-node and evolve into C-character with the
probability $w_{ij}$ in the {\em same} time step.  Thus, the key
features for a last surviving pattern are: (i) we need at least
two connected D-nodes, and (ii) the C-nodes that are connected to
only one of these surviving D-nodes have a higher chance to
replace the D-nodes.  The discussion also serves to point out the
importance of run time in numerical studies. When the system
evolves to the last surviving pattern, the pattern may or may not
evolve into an all-C pattern.  It only happens probabilistically.
The time it takes, even when $r < r_{D}$, for the all-C pattern
to emerge depends on the values of $r$ and $K$.  For $r
\rightarrow r_{D}$ and for larger values of $K$, the time may be
exceedingly long so that a smaller value of $r_{D}$ than that
theoretically allowed will be determined numerically.

\begin{figure}
\begin{center}
\epsfig{figure=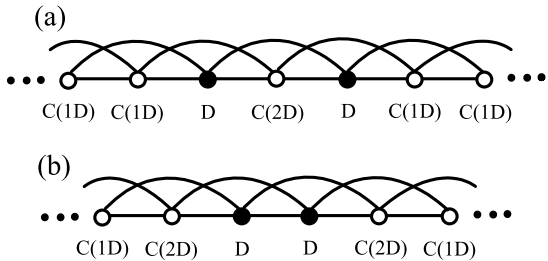,width=0.95\linewidth} \caption{Two last
surviving patterns in $K=2$ regular lattices for estimating the
extinction payoff $r_{D}$.  Typically, these patterns consist of
two connected D-nodes in a background of C-nodes.  The C-nodes
connected to one D-node (labelled C(1D)) and connected to two
D-nodes (labelled C(2D)) have different average payoffs.  $r_{D}$
is found by equating the average payoffs $\overline{V_{C(1D)}}$
and $\overline{V_{D}}$.}
\end{center}
\end{figure}

Extending the argument to $K=2$ regular lattices, a single D-node
will evolve into one of the last surviving patterns shown in
Figure 2.  In both patterns, there are two types of neighboring
C-nodes.  For the C-nodes connected to only one D-node, the
average payoff $\overline{V_{C(1D)}} = 1 - r/4$.  For the C-nodes
connected to two D-nodes, $\overline{V_{C(2D)}} = 1 - r/2$.  For
the D-nodes, $\overline{V_{D}} = 3(1+r)/4$.  Note that in general
$\overline{V_{C(1D)}} > \overline{V_{C(2D)}}$.  Since there are
more C-nodes with one D-neighbor in the pattern in Figure 2(a)
than Figure 2(b), the D-nodes in Figure 2(a) are most likely to be
the last surviving pattern in numerical simulations.  There are
two values of $r$ that are of interest.  For $r < 1/5 = 1/(2K+1)$,
we have $\overline{V_{C(1D)}} > \overline{V_{C(2D)}} >
\overline{V_{D}}$. Therefore, all the neighboring C-nodes have a
finite probability to replace the two D-nodes in one time step.
This is why it is easier to observe an all-C pattern for $r <
1/5$. In principle, $r_{D}$ can be found by equating
$\overline{V_{C(1D)}}$ and $\overline{V_{D}}$, giving $r_{D}=1/4$.
However, for $1/5 < r < 1/4$, we have $\overline{V_{C(1D)}} >
\overline{V_{D}} > \overline{V_{C(2D)}}$.  These inequalities
imply that even if we arrive at the patterns in Figure 2, the
chance of replacing {\em both} D-nodes by C-character is small,
since only the C-nodes connected to one D-node can carry out the
replacement. Instead, the D-nodes may replace the C-nodes linked
to both D-nodes to arrive at a pattern with more D-nodes.  This
prevents the all-C pattern from emerging in any run of reasonably
long run time. This effect is obviously more severe as $K$
increases, as there will be more C-nodes with two D-neighbors.

Generalizing the argument to arbitrary value of $K$, we have
$\overline{V_{D}} = (1+r)(1 - 1/2K)$ and $\overline{V_{C(1D)}} = 1
- r/2K$.  Thus, {\em in principle}, $r_{D}(K,q=0) = 1/2K$.
However, for any practical run times, the all-C pattern is hard to
observe for $1/(2K+1) < r < 1/(2K)$, except for $K=1$ and $2$.
Thus, the numerical results in Figure 1, which are obtained by
fixing the run time to be $t=10^{4}$ time steps, give $r_{D}
\approx 1/(2K+1)$ for $K \geq 3$. We have tried longer run times,
e.g., $3 \times 10^{5}$ time steps, and the value of $r_{D}$ only
tends to approach the value $1/2K$ very slowly.

\section{Results: Watts-Strogatz re-wired networks}

\begin{figure}
\begin{center}
\epsfig{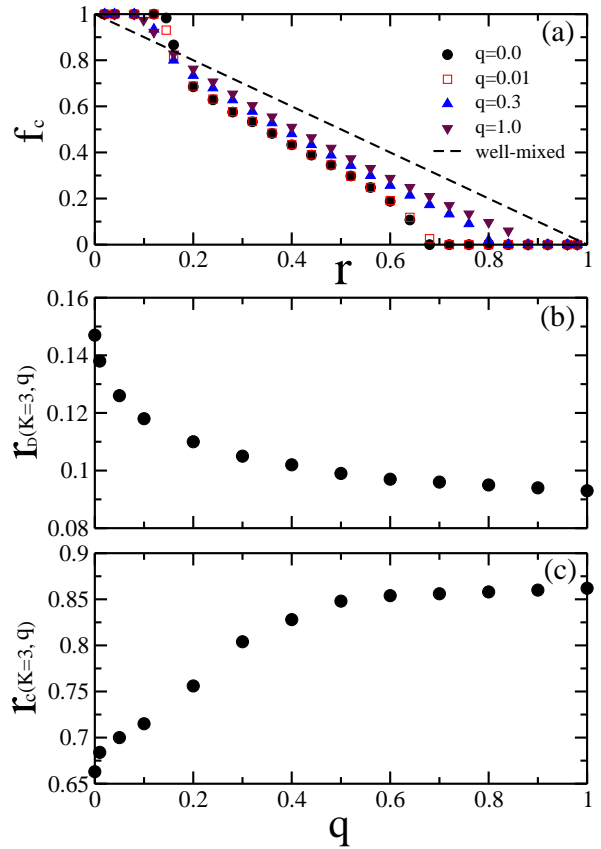} \caption{(Color
online) (a) The frequency of cooperators $f_{C}$ as a function of
the the payoff parameter $r$ in Watts-Strogatz networks of $K=3$
for different values of the re-wiring probability $q=0$, $0.01$,
$0.3$, $1.0$.  The dashed line indicates the result $f_{C}=1-r$
for a fully connected network. (b) The extinction payoff
$r_{D}(K,q)$ as a function of $q$ for $K=3$. (c) The extinction
payoff $r_{C}(K,q)$ as a function of $q$ for $K=3$.}
\end{center}
\end{figure}

Going beyond regular lattices, Figure 3(a) shows $f_{C}$ as a
function of $r$ in Watts-Strogatz (WS) networks with different
re-wiring probabilities $q=0.01$, $0.3$ and $1.0$. The network
consists of $N=10^{3}$ nodes and $K=3$, with the run time fixed at
$t=10^{4}$ time steps.  Results in a regular lattice ($q=0$) and
fully connected networks (dashed line) are also included for
comparison. It is noted that the effect of increasing $q$ for
fixed $K$ is similar to increasing $K$ for fixed $q=0$ (regular
lattices).  The main features are that as $q$ increases, the
extinction payoff $r_{D}$ decreases and the $f_{C}$ is higher than
that in a regular lattice with the same value of $K$ for a large
range of $r$. Figures 3(b) and (c) show that, for fixed $K=3$, the
extinction payoff $r_{D}$ ($r_{C}$) drops (rises) with the
re-wiring probability $q$. We have checked that while the shortest
path $L(q)$ changes sensitively with $q$, it is however not the
determining factor for the extinction payoffs.  Will the
clustering coefficient $C$ matter? For regular lattices, $C$
increases with $K$ and saturates at large $K$ following
$C(0)=3(2K-2)/[4(2K-1)]$ \cite{watts,albert}. If the drop in
$r_{D}$ with $K$ in regular lattices were attributed to the
increase in $C$, then one would have expected $r_{D}$ to {\em
increase} with $q$ as $C(q) \simeq C(0)(1-q)^{3}$ decreases with
$q$ in WS networks \cite{watts,albert}. This is, however, {\em
not} what is observed in numerical results. Therefore, the
clustering coefficient $C$ is also not a determining factor for
$r_{D}$ and $r_{C}$.  These observations are in sharp contrast to
the networking effects in evolutionary PD games
\cite{abramson,kim,ebel,masuda,wu}.  Noting that the WS networks
have a fixed $\langle k \rangle =2K$ with or without re-wiring,
$r_{D}$ and $r_{C}$ cannot be determined by the mean degree
$\langle k \rangle$.  Thus, $r_{D}$ and $r_{C}$ are not determined
by the commonly studied quantities in networks.
\begin{figure}
\begin{center}
\epsfig{figure=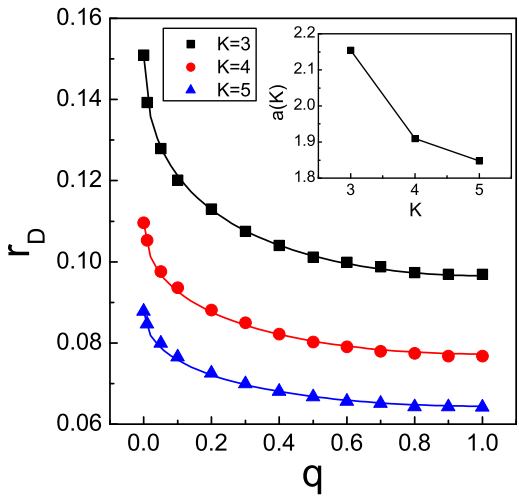,width=0.8\linewidth} \caption{(Color
online) The extinction payoff $r_{D}(K,q)$ is plotted against $q$
for different values of $K=3$, $4$, $5$.  The results are obtained
by runs of $3 \times 10^{5}$ time steps.  The lines are obtained
by fitting to a form given by Eq.(4).  The inset gives the
dependence of the fitting parameter $a(K)$ on $K$.}
\end{center}
\end{figure}

\begin{figure}
\begin{center}
\epsfig{figure=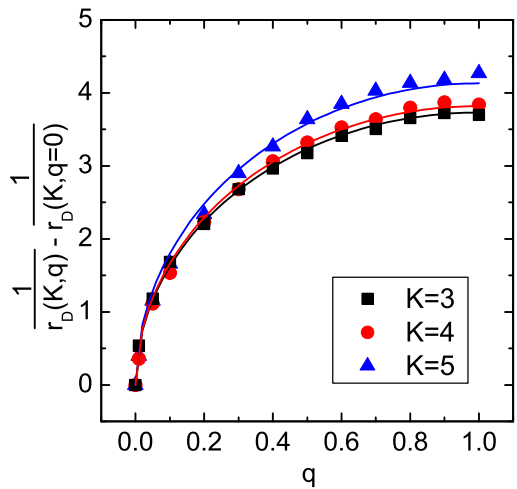,width=0.8\linewidth} \caption{(Color
online) By suitably choosing the quantity on the y-axis, the set
of scattered data in Fig. 4 can be shown to follow a similar
functional dependence on $q$.  The dependence is found to follow
that of $\sqrt{2q - q^{2}}$ in the standard deviation of the
degree distribution in Watts-Strogatz networks after re-wiring for
fixed $K$.  The lines are obtained by using the fitting parameters
obtained in Fig. 4.}
\end{center}
\end{figure}

The results in Figures 1 and 3 revealed that the extinction
payoffs $r_{D}(K,q)$ and $r_{C}(K,q)$ depend on both the
structural parameters $K$ and $q$.  It does remain an intriguing
problem of which geometrical property (properties) of a re-wired
network determines $r_{D}$ and $r_{C}$ in the Watts-Strogatz
networks.  In what follows, we will focus on analyzing
$r_{D}(K,q)$, as $r_{C} \approx 1 - 2 r_{D}$. Figure 4 shows
collectively the numerical results (symbols) of $r_{D}(K,q)$ for
different values of $K$ and $q$, for a longer fixed run time of $3
\times 10^{5}$ time steps. As discussed, the numerical results for
$r_{D}(K,q=0)$ in a regular lattice may not come out to be $1/2K$
even using a longer run time, we could avoid this problem and
focus on the effects of re-wiring, i.e., $q \neq 0$, by exploring
the behavior of the difference in $r_{D}(K,q)$ and $r_{D}(K,0)$.
Interestingly, we found that by plotting $(1/r_{D}(K,q) -
1/r_{D}(K,q=0))$ as a function of $q$, the numerical data for
different values of $K$ basically follow the same functional form,
as shown by the symbols in Figure 5.  Examining the
$q$-dependence, we notice that the behavior is similar to that in
the {\em standard deviation} $\sigma(K,q)$ of the degree
distribution $P(k)$ in the Watts-Strogatz networks.  The re-wiring
process changes $P(k)$ gradually from a delta-function at $k=2K$
for $q=0$ to a distribution that has a lower cutoff at $k=K$ with
a finite $\sigma(K,q)$ that increases with $q$ \cite{albert}. More
specifically,
\begin{equation}
\sigma(K,q) = \sqrt{K} \cdot \sqrt{2q-q^{2}},
\end{equation}
which follows from the degree distribution of the Watts-Strogatz
model.  Motivated by this observation, we use the numerical data
in Figure 4 and perform a fit to the functional form
\begin{equation}
r_{D}(K,q) = \frac{1}{\frac{1}{r_{D}(K,0)} + a(K) \sigma(K,q)},
\end{equation}
where $a(K)$ is a fitting parameter.  The first term in the
denominator of Eq. (4) imposes the restriction that the fitted
line must pass through the $q=0$ data point.  The fitted lines are
also shown in Figure 4.  With only one fitting parameter for each
value of $K$, the lines fit the data accurately. This implies that
the $q$-dependence follows that in $\sigma(K,q)$. The fitting
parameters are found to be $a(3)=2.154$, $a(4)=1.910$, and
$a(5)=1.847$. Note that $a(K)$ drops with $K$ (see inset in Fig.
4). Since $\sigma(K,q)$ carries a factor of $\sqrt{K}$, the
combination $b(K) = a(K) \sqrt{K}$ becomes quite insensitive to
$K$, with $b$ increasing from $3.73$ for $K=3$ to $4.13$ for
$K=5$. As a consistency check, we also plot the fitted lines in
Figure 5, using the same coefficients $a(K)$ as obtained in Figure
4. Note that even for $q=1$, the spread in the data and fitted
lines is small, reflecting the weak $K$-dependence in the
coefficients $b(K)$. The extinction payoff is thus found to be
described by
\begin{equation}
\frac{1}{r_{D}(K,q)} = \frac{1}{r_{D}(K,q=0)} +  b(K) \; \sqrt{2q
- q^{2}},
\end{equation}
where $b$ carries only a weak $K$-dependence.  Our results on the
extinction payoff $r_{D}(K,q)$ in the Watts-Strogatz networks can
be summarized as follows. The dependence on $K$ is dominated by
that in $r_{D}(K,q=0)$ and the $q$-dependence follows that of the
standard deviation $\sigma(K,q)$ of the degree distribution.

\section{Discussion}

Our analytic approach in getting $r_{D}$ in regular lattices (see
Sec. III) can also be applied to re-wired networks.  The
difference is that in a re-wired network, different nodes may have
different number of neighbors, i.e., different degrees.  Again, we
consider a last surviving pattern with two connected D-nodes with
otherwise C-neighbors.  Let $k_{D}$ be the degree of one of the
D-nodes. Therefore, this D-node has $(k_{D}-1)$ C-neighbors and
one D-neighbor.  Let $k_{C}$ be the degree of one of these
C-neighbors or the typical degree of these C-neighbors.  We assume
that the C-nodes are only connected to one D-node, as they are the
ones more likely to replace the D-nodes.  For the D-node, the
average payoff is:
\begin{equation}
\overline{V_{D}} = \frac{k_{D}-1}{k_{D}} +
\frac{(k_{D}-1)r}{k_{D}},
\end{equation}
while for the C-node, the average payoff is:
\begin{equation}
\overline{V_{C}} = 1 - \frac{r}{k_{C}}.
\end{equation}
For regular lattices, $k_{C} = k_{D} = 2K$ and $r_{D}=1/2K$ is
recovered when we equate $\overline{V_{D}}$ to $\overline{V_{C}}$.
In addition, as $K$ increases in a regular lattice, $r_{D}
\rightarrow 0$, which is the result in the well-mixed case.  In
general, equating Eq. (6) and Eq. (7) gives
\begin{equation}
r'_{D} = \frac{k_{C}}{(k_{D}-1)k_{C} + k_{D}}.
\end{equation}
The meaning of $r'_{D}$ is that, if a last surviving pattern of
the considered structure is approached in a numerical simulation,
then for $r < r'_{D}$, such a pattern may become all-C in one time
step.

Equation (8) suggests several possibilities on estimating $r_{D}$.
A lower bound of $r_{D}$ can be constructed by noting that a
D-node occupying a node of high degree can take advantage of the
many C-neighbors and hence requires a lower value of $r$ to
replace it.  Similarly, a surrounding C-node with fewer
C-neighbors has lower average payoff and hence is harder to
replace the D-nodes.  Let $k_{max}(K,q)$ be the maximum degree in
a re-wired network for given $K$ and $q$.  The minimum degree in a
Watts-Strogatz network is close to $K$ after re-wiring.  An
extreme (but rare) case is that of a D-node occupying a degree
with $k_{max}$ being surrounded by C-nodes that occupy nodes with
degree $K$. A more reasonable assumption is to take the degree of
the C-nodes as the mean degree $2K$, as any one of the C-nodes
may be picked in the evolution step.  We can substitute $k_{D} =
k_{max}(K,q)$ and $k_{C}=2K$ to get an estimate of a lower bound
\begin{equation}
r_{D}^{lower} = \frac{2K}{2K(k_{max}(K,q)-1)+ k_{max}(K,q)}.
\end{equation}
We have checked that Eq. (9) indeed gives values that are lower
than numerical results.  More interestingly, $k_{max}(K,q)$ turns
out to depend on $q$ through the standard deviation $\sigma(K,q)$
\cite{remark1} . Hence, Eq. (9) does reproduce the drop of $r_{D}$
as $q$ increases, as observed in Fig.4.

It is also possible to relate Eq. (8) to Eqs. (4) and (5).
Assuming that, after averaging over runs in numerical studies,
that $k_{C}$ and $k_{D}$ can both be approximated as some typical
degree $k_{ave}(K,q)$, then Eq. (8) suggests that the extinction
payoff would be
\begin{equation}
r_{D} = \frac{1}{k_{ave}(K,q)}.
\end{equation}
Since the dynamics in evolutionary SG depends on the average
payoffs, which in turn depends on the geometrical structure of the
neighborhood of a node, we expect in general $k_{ave}$ is
different from the mean degree $2K$.  Instead, D-node tends to
survive easily by staying on nodes of higher degree.  Comparing
Eq.(10) with Eq. (4), we notice that the fitted result to
numerical data implies that
\begin{equation}
k_{ave}(K,q) = \frac{1}{r_{D}(K,q=0)} + a(K) \sigma(K,q).
\end{equation}
where the first term should in principle be $2K$.

In summary, we investigated the extent of cooperation that would
emerge in a networked evolutionary snowdrift game.  The random
re-wiring model of Watts and Strogatz is used.  Comparing to a
fully connected network, a spatial structure of regular lattices
suppresses $f_{C}$ over a wide range of the payoff $r$ and
re-wiring lowers the suppression.  We identified two extinction
payoffs $r_{D}$ and $r_{C}$.  For regular lattices, $r_{D}$
should take on the value of $1/2K$, although a value closer to
$1/(2K+1)$ is usually observed in numerical studies.  The
dependence of $r_{D}(K,q)$ on $K$ and $q$ is highly non-trivial.
The key network property that gives the $q$-dependence is found
to be the standard deviation of the degree distribution. This
finding, in turn, implies that it is the existence of nodes with
higher degrees due to randomly re-wiring that plays a dominant
role in determining the extinction payoffs.

\acknowledgments{We thank P.P. Li of CUHK for useful discussions.
One of us (P.M.H.) acknowledges the support from the Research
Grants Council of the Hong Kong SAR Government under Grant No.
CUHK-401005. The work was completed during a visit of D.F.Z.,
L.X.Z. to CUHK which was supported by a Direct Grant of Research
from CUHK.  This work was also supported in part by the National
Natural Science Foundation of China under Grant Nos. 70471081,
70371069, and 10325520, and by the Scientific Research Foundation
for the Returned Overseas Chinese Scholars, State Education
Ministry.}


\begin{thebibliography}{99}
\bibitem{remark} This is the case except for a few isolated values of
$r$ with $r > r_{C}$ for which the chain can be evolved into a
configuration with a majority of D-nodes with isolated C-nodes
distributed at equal distances.
\bibitem{remark1} We have checked that the maximum degree for
given $K$ and $q$ takes on the form $k_{max}(K,q) = 2K + c(K)
\sigma(K,q)$, where $c(K)$ depends on $K$ only weakly and takes on
values close to $4$ for different values of $K$.  As $\sigma(K,q)$
increases with $q$, $r_{D}^{lower}$ in Eq. (9) drops with $q$, as
observed in numerical results.
\end{thebibliography}
\end{document}